**Health System Scale Semantic Search Across Unstructured Clinical Notes**

Faith Wavinya Mutinda[1], Spandana Makeneni[1], Anna Lin[1], Shivaji Dutta[2], Irit R. Rasooly[3], Patrick Dibussolo[1], Shivani Kamath Belman[1], Hessam Shahriari[1], Kevin Murphy[1], Alex B. Ruan[1,4], Barbara H. Chaiyachati[3], Sanjay Chainani[3], Robert W. Grundmeier[1,3], Scott M. Haag[1], Jeffrey M. Miller[1], Heather M. Griffis[1], Ian M. Campbell[1,3,5,*]

1) Department of Biomedical and Health Informatics, Children's Hospital of Philadelphia, Philadelphia, PA
2) Google Cloud, Sunnyvale, CA
3) Department of Pediatrics, University of Pennsylvania, Philadelphia, PA
4) Division of Neonatology, Children's Hospital of Philadelphia, Philadelphia, PA
5) Division of Human Genetics, Children's Hospital of Philadelphia, Philadelphia, PA
* Correspondence to: Ian M. Campbell (campbellim@chop.edu)

**Abstract**

**Introduction:** Semantic search, which retrieves documents based on conceptual similarity rather than keywords, offers substantial advantages for retrieval of clinical information. However, deploying semantic search across entire health systems, comprising hundreds of millions of clinical notes, presents formidable engineering, cost, and governance challenges that have prevented institutional adoption at scale.
**Methods:** We deployed a semantic search system at a large children's hospital indexing 166 million clinical notes (484 million embedding vectors) from 1.68 million patients. The system uses instruction-tuned qwen3-embedding-0.6B embeddings, stores vectors in a managed database with storage-optimized indexing, maintains full-text metadata in a low-latency key-value store, and operates within a HIPAA-compliant governance framework. We evaluated the system by optimizing the embedding model and chunking strategy using a physician-authored benchmark dataset, characterizing full-scale performance (cost, latency, retrieval quality), and assessing clinical utility via chart abstraction efficiency and comparison to ICD-10 cohort generation.
**Results:** The system delivers sub-second query latency with monthly operational costs of approximately USD 4,000. Qwen3 embeddings with 300-token chunk size achieved 94.6% accuracy on a clinical question-answering benchmark. In clinical utility evaluation across three abstraction tasks, semantic search reduced time-to-completion by 24 to 89% versus keyword-based chart review while maintaining inter-rater agreement where assessable. During system-wide cohort retrieval, semantic search recovered 98% of patients with molecularly confirmed genetic diseases, versus at most 75% by diagnosis codes.
**Conclusion:** Health-system-scale semantic search is both technically and operationally feasible. The system provides institutional infrastructure supporting interactive search, cohort generation, and downstream LLM-powered clinical applications without requiring specialized informatics expertise.

## Introduction

Much of what matters in a patient's medical record, including clinical reasoning, differential diagnoses, family history, and imaging impressions, is written in free text. While structured electronic health records (EHR) fields capture discrete data points, unstructured notes remain the primary medium through which clinicians document. Accessing this information at scale has been a persistent challenge: keyword search fails on synonyms, abbreviations, negations, and paraphrases, and direct database queries typically require specialized informatics support, making large-scale access difficult for nonexpert clinical researchers[1].

Semantic vector search offers a fundamentally different approach. Instead of matching exact terms, it encodes text into dense vector representations that capture meaning, enabling retrieval based on conceptual similarity rather than lexical overlap. When combined with retrieval-augmented generation (RAG)[2], semantic search can provide large language models (LLMs)[3] with patient-specific context they otherwise cannot access, improving the factual grounding of generated outputs and reducing hallucinations.

Despite growing interest, prior healthcare semantic search efforts have largely concentrated on narrower use cases such as literature retrieval or task-specific cohorts rather than health-system-wide search across all clinical notes[4,5]. Prior efforts target settings such as radiology reports, clinical guidelines, or specialty-specific applications[6–12]. These efforts do not address the central engineering challenge: *can semantic search operate across an entire health system's clinical notes, with the cost, latency, and governance properties required for sustained institutional deployment?*

We sought to address this question directly. This manuscript describes the design, implementation, and evaluation of a semantic search system deployed at the Children's Hospital of Philadelphia (CHOP) that indexes 166 million clinical notes for 1.68 million patients as 484 million embedding vectors. The system delivers sub-second query latency at a low monthly operational cost, supports fine-grained metadata filtering, and operates within a HIPAA-compliant governance framework with project-level access controls and audit logging.

The system was evaluated across three experiments: retrieval quality using a physician-authored clinical question-answering benchmark, system performance under concurrent workload, and clinical utility through a controlled comparison of chart abstraction efficiency. The system is designed not as a point solution but as institutional infrastructure: a shared semantic index over which interactive search, cohort generation, and downstream LLM-powered applications can be built.

## Methods

Our prospective evaluation of a semantic search system was conducted in the context of a multi-specialty pediatric healthcare setting. Our study had 3 phases as detailed below: (1) evaluation and selection of embedding model and chunking strategy; (2) system performance and cost; and (3) accuracy and efficiency of the system in the context of a clinical chart review task.

### Setting

The Children's Hospital of Philadelphia (CHOP) is a multi-specialty quaternary pediatric healthcare system caring for 400,000 patients annually across two hospitals. This study included patients cared for at CHOP from January 2000 through September 2025 and was approved by the CHOP Institutional Review Board (IRB #22-020587) as well as the CHOP AI Governance Committee.

### Data

The CHOP EHR contained over 228 million clinical notes. After excluding notes without text content and those containing specially protected health information, 166 million notes for 1.68 million patients were indexed. The notes span over 80 specialties and were authored by over 44,000 providers. Associated metadata for each note includes patient demographics (name, date of birth, sex), note attributes (type, department, specialty), author information (name, role), and timestamps.

### System architecture

The semantic system comprises four main components: a text processing pipeline, a vector database, a key-value metadata store, and a web-based user interface (Figure 1). Clinical notes are extracted from the CHOP EHR, split into overlapping chunks, and encoded into dense vectors using an embedding model. The vectors are indexed in a managed vector database that supports approximate nearest-neighbor retrieval and filtering on structured metadata fields. Full note text and detailed metadata are stored separately in a low-latency key-value store to reduce storage overhead. A web-based user interface combines natural-language querying with metadata filters for interactive search (Figure 2). All components operate within a HIPAA-compliant environment with access controls and audit logging.

### Chunking & embedding

Clinical notes frequently contain multiple distinct concepts (chief complaint, history, medications, results, assessment and plan) within a single document. A single embedding for an entire note dilutes the representation of any individual concept. Chunking creates semantically coherent segments that allow the retrieval system to surface the most relevant context. We evaluated chunk sizes ranging from 200 to 600 tokens with a 50-token overlap and selected a 300-token chunk size as the production configuration (see Results). The token overlap preserves concepts that span chunk boundaries. Additionally, the chunking pipeline prioritizes natural linguistic boundaries (paragraph breaks, line breaks) to maintain coherence and avoid fragmenting clinical concepts across chunks.

To encode the resulting chunks into a dense vector space for semantic retrieval, we evaluated four embedding models: qwen3-embedding-0.6B[13], nomic embed[14], E5 large[15], and BioClinical BERT[16]. Models with output dimensionality at or below 1024 were prioritized to limit cloud storage costs. Based on benchmarking results (see Results), qwen3-embedding-0.6B was selected as the production embedding model. Qwen3-embedding-0.6B is an instruction-tuned embedding model that produces 1024-dimensional vectors and accepts task instructions that tailor the embedding to the retrieval context; at query time, each query is prefixed with a retrieval instruction. Embeddings were computed using last-token pooling[17] and L2 normalized to unit length so that dot-product similarity reflects directional alignment rather than magnitude.

Embedding computations were performed on Google Cloud Trillium TPUs using PyTorch/XLA. The corpus was partitioned by year and month, with each partition processed independently. Chunking the 166 million notes resulted in 484 million chunks, which were embedded over approximately 11 calendar days using TPU allocations provided through Google's TPU Research Cloud (TRC) program.

### Vector database indexing

Vectors were stored in Google Cloud Vector Search, a managed vector database that supports both in-memory (RAM-based) and storage-optimized (disk-based) deployment tiers. We selected the storage-optimized tier, which trades modestly higher retrieval latency for substantially lower hosting costs[18]. Retrieval uses dot-product similarity via ScaNN (Scalable Nearest Neighbors) and SOAR[19]. ScaNN employs hierarchical clustering and asymmetric hashing to narrow the search space efficiently, then rescores approximate candidates at full precision[19].

The index supports query-time filtering on categorical attributes (patient identifier, note type, encounter type, department, specialty, author type, author name) and numeric ranges (date, age) (Supplementary Table 1). These filters both improve retrieval precision and reduce noise. This is particularly important because dense similarity scoring tends to favor short, narrowly focused documents whose embeddings are tightly aligned with a query. Filtering by note category allows users to prioritize clinically substantive documents.

The vector database supports incremental insertion without requiring a full index rebuild. We have designed a pipeline to extract, chunk, embed, and insert new notes on a regular schedule. Full automation of this pipeline, including scheduling and monitoring, is ongoing work.

### Metadata & note text serving

The vector database stores only note identifiers, embeddings, and filterable metadata. Because vector database pricing scales with the total size of stored data, excluding full note text and detailed metadata substantially reduces hosting costs. Full note text and metadata are instead maintained in a low-latency key-value store (Bigtable), which offers significantly lower storage costs. Row keys use a salt-and-reverse scheme to prevent hot spotting on sequential note IDs. At query time, the top-k note identifiers returned by vector search are used to fetch the corresponding full records from the key-value store with minimal added latency.

**Benchmarking**

To evaluate retrieval quality, we created a patient-specific multiple-choice question-answering benchmarking dataset (CHOP_MCQA_v0.5). Three physicians (I.R.R, A.B.R, and I.M.C) authored 334 questions tied to 322 unique patients, spanning multiple specialties including genetics, cardiology, rheumatology, social work, and neonatology. Each question has one correct answer and four plausible distractors designed to simulate the ambiguity of real clinical information needs (Supplementary Table 2). Questions spanned both temporal and time-independent information needs: 108 of the 334 questions (32%) referenced a specific calendar year and thus required retrieval of time-scoped evidence, while the remainder were not tied to a particular date.

All the embedding models and chunking configurations were evaluated on a reduced index containing only the notes for benchmark patients, as evaluations over the full corpus would have incurred prohibitive cost. Each question was embedded, the top-20 most relevant chunks retrieved (filtered to the target patient), and an LLM, DeepSeek-R1-Distill-Llama-70B[20], assessed whether it could select the correct answer using the retrieved context. Each question was evaluated across five independent runs with stochastic decoding, and the final answer was determined by majority vote.

Because ground-truth note-level relevance labels are unavailable and would require costly and time-intensive annotation, we used end-to-end accuracy as a pragmatic proxy for retrieval quality. A correct answer implies that retrieval surfaced sufficient evidence, and the LLM interpreted it correctly; an incorrect answer may reflect failure in either stage. Because the LLM was held constant across all conditions, differences in accuracy isolate the effect of embedding model and chunk size on retrieval. To test whether the best-performing configuration (qwen3-embedding-0.6B at 300 tokens) differed significantly from each alternative embedding model at the same chunk size, we applied McNemar's test to paired per-question outcomes (correct vs. incorrect), which is appropriate because the same 334 questions were evaluated under every configuration; a Bonferroni-corrected significance threshold of $p < 0.0167$ was applied across the three pairwise comparisons.

To evaluate retrieval quality independently of the language model, we obtained human relevance judgments for the chunks retrieved by the production configuration. For each of the 334 benchmark questions, the top-20 chunks returned by the production system (qwen3-embedding-0.6B, 300-token chunks), filtered to the target patient, were presented to a physician annotator. Each chunk was labeled against a binary criterion: could a clinician, given only that chunk and the question stem, select the correct answer over the four distractors? Chunks were labeled in rank order to a depth of 10 for every question; for questions with no answering chunk in the top 10, annotation was extended to rank 20. From these labels we computed Hit Rate@k (the fraction of questions with at least one answering chunk in the top k), precision@k (the mean fraction of top-k chunks judged to answer the question), and mean reciprocal rank (MRR), overall and stratified by answer polarity, each with 95% Wilson score confidence intervals. Inter-rater reliability was assessed on a polarity-stratified 50-question subset labeled independently by a second physician using Cohen's $\kappa$[21].

## Clinical utility

To assess the practical utility of the semantic search system for clinical research, we asked five clinicians (A.L., I.R.R., B.H.C., S.C., and I.M.C) to complete three abstraction tasks reflecting real research workflows: (1) identification of documented genetic conditions, (2) documentation of age at first lifetime seizure, and (3) cohort generation for ballet-related foot injuries. We compared two search methods: traditional EHR abstraction and the proposed semantic search system.

Tasks 1 and 2 were designed to model structured abstraction scenarios in which the patient cohort is fixed, and the objective is to extract specific clinical data from the chart. For each task, a cohort of 20 patients was selected. Each abstractor reviewed all 20 patients, with half of the patients assigned to traditional EHR extraction and half to semantic search. Method assignment was randomized independently per abstractor. Task 3 models cohort discovery, where eligible patients must be identified from medical records. For the EHR method, abstractors reviewed 100 candidates retrieved via a substring query against the full clinical note corpus: a note qualified if its lowercased text contained the term "ballet" and at least one anatomical site term ("foot," "ankle," or "toe") and at least one injury term ("injury," "fracture," "sprain," or "strain"). For semantic search, abstractors used the system alone to discover qualifying patients de novo from the full 1.68-million-patient corpus, without a pre-filtered candidate list.

For the traditional EHR abstraction condition, each abstractor used their personal Epic Hyperspace instance without standardized instructions. This design was intentional: rather than prescribing a uniform workflow, we sought to maximize ecological validity by allowing abstractors to employ the tools and habits they use in actual research practice. Hyperspace offers a range of chart navigation and search capabilities, such as “Chart Search”, which supports free-text queries across note content, as well as note-type and date filters; however, familiarity with and use of these features varied across abstractors as it does in routine clinical research. No constraining protocol was imposed on the EHR condition because naturalistic variation in search behavior is inherent to the real-world comparator this study aims to benchmark against.

The primary outcome for all tasks was time-to-completion per patient, compared between methods using the Mann-Whitney U test[22]. For tasks 1 and 2, we also assessed whether the two methods yielded equivalent extracted information through inter-rater reliability analysis. For each task, the effect size of the time-to-completion difference was quantified as the rank-biserial correlation (r). Sample sizes (20 patients per task for tasks 1 and 2; 100 candidates for task 3) were determined pragmatically by annotator availability and the exploratory nature of the evaluation rather than by a formal a priori power calculation; the large effect sizes observed across all three tasks ($r = 0.29$, 0.47, and 0.83 for tasks 1, 2, and 3, respectively) indicate that the study was adequately sensitive to detect the time-to-completion differences reported. The inter-rater agreement was calculated overall and separately for within-method pairs (EHR vs EHR) and cross-method pairs (EHR vs semantic). For the categorical outcome (task 1), overall agreement was quantified using Fleiss' κ[23] and pairwise within- and cross-method agreement using Cohen's κ[21]. For the continuous outcome (task 2), both overall and pairwise agreement were quantified using Krippendorff's α[24]. To test whether agreement differed by search method, a patient-level nonparametric bootstrap (10,000 resamples) was used to estimate the difference between within-method and cross-method agreement for each task. A p-value was derived from the proportion of bootstrap resamples in which the difference was as large or larger than the observed difference.

To assess whether faster cohort discovery in task 3 came at a cost to accuracy, we estimated the precision of each arm's selected patients. A selected patient was counted as a true positive if independently selected by two or more abstractors; each singleton selection was confirmed by a second clinician blinded to ascertainment arm. Precision was calculated as the fraction of an arm's uniquely selected patients confirmed eligible, with 95% Wilson score confidence intervals. Because the two arms operated on disjoint candidate pools of different sizes, precision was not compared across arms.

**Cross-patient genetic disease retrieval**

To evaluate retrieval across the entire patient population rather than within a single chart, we measured the system's ability to recover molecularly defined disease cohorts from the full 1.68-million-patient index. Ground-truth cohorts were derived from Epic "Genomic Indicators," structured flags applied during routine clinical care to patients with a molecularly confirmed genetic diagnosis. Because indicators are assigned on the basis of molecular test results rather than diagnosis coding, they identify confirmed patients independently of whether any corresponding coded encounter ever occurred, providing a coding-independent reference standard. We evaluated all 71 conditions represented in the index (812 confirmed patients).

For each condition, a single natural-language query describing the molecular diagnosis (Supplementary Table 3) was issued against the full, unfiltered index with no patient or metadata filter. Retrieved notes were mapped to patients, patients were ranked by their maximum chunk similarity score, and patient-level recall@k was computed as the fraction of confirmed patients appearing within the top-k unique patients.

As a structured-query comparator, for each condition, we selected the most specific ICD-10-CM diagnosis code naming the condition or its principal phenotype, together with a genetic-susceptibility or family-history code where an applicable one existed (Supplementary Table 3). The selected codes were expanded to all descendant codes. For 29 of the 71 conditions, no ICD-10-CM code names the syndrome, and the closest available surrogate was used. For each method we also report the recall ceiling, the maximum recall achievable at unlimited retrieval depth; for the structured comparator this equals the fraction of confirmed patients with any relevant coded encounter and is therefore an upper bound on recall by any ICD-based query at any depth.

Precision was not estimable against this reference standard: the Genomic-Indicator registry is an incomplete, institution-specific enumeration, so a returned patient absent from the registry may represent a genuine but un-annotated confirmed patient rather than a false positive. We therefore report recall only.

**Security and data governance**

The system operates within Arcus[25], CHOP's institutional research platform hosted on a HIPAA-compliant Google Cloud environment under a Business Associate Agreement. Access is governed at the project level: each institutional review board (IRB)-approved research project receives its own containerized deployment, authenticated to a project-specific data environment. An allowlist restricts each deployment to only the note identifiers the project has been approved to access, and retrieved notes are filtered against this allowlist before display. Specially protected health information (e.g., substance abuse treatment, psychotherapy notes) is excluded upstream by the data delivery team.

All searches, including query text, filters, user identity, and returned note identifiers, are logged for auditing and compliance monitoring. We partnered with the IRB to develop a standardized protocol attachment that investigators append to their IRB applications when requesting access to the system. This attachment describes the system's operation, the scope of access requested (patient-specific or health-system-wide), and the associated risks.

## Results

### Embedding model and chunking strategy

We evaluated all the embedding models and chunking configurations on the CHOP_MCQA_v0.5 dataset. Figure 3 summarizes the results. Qwen3-embedding-0.6B achieved the highest accuracy across all chunk sizes, with the 300-token configuration yielding 95.5% (95% Wilson CI: 92.7%--97.3%). McNemar's tests on the paired per-question outcomes confirmed that qwen3-embedding-0.6B at 300 tokens significantly outperformed BioClinical BERT ($p = 1.97 \times 10^{-9}$) and E5 large ($p = 0.002$), while its advantage over nomic embed did not reach significance after correction ($p = 0.029$). Three of the four models surpassed 90% accuracy, indicating that with an appropriate embedding model, retrieved notes provide sufficient context for accurate clinical question-answering. At the time of evaluation, qwen3-embedding-0.6B was among the top-performing models on the multilingual text embedding benchmark (MTEB) leaderboard[26] for its dimensionality. BioClinical BERT, despite being pre-trained on clinical text, had the lowest accuracy. However, this model is also substantially smaller (~110M vs. 600M parameters), uses a different architecture, and lacks instruction tuning; the comparison does not isolate the effect of domain-specific pretraining. Qwen3-embedding-0.6B with 300-token chunks size was selected as the production configuration. Accuracy increased with retrieval depth and plateaued near $k = 20$ retrieved notes (Supplementary Figure 1), supporting the use of this retrieval threshold for all subsequent evaluations.

Error analysis on the best-performing configuration revealed that incorrect answers were mainly associated with the absence of relevant documents in the top 20 retrieved results, particularly for queries requiring temporal reasoning. Temporal mismatches occurred when questions referred to specific years, but the retrieval system returned documents from different time periods. For example, for a question about the reason for a 2020 hospitalization, the retrieval system returned records from 2022 to 2024, preventing the model from identifying the correct answer. This limitation arose because temporal filters were not applied in the benchmarking setup. This failure mode can be mitigated by applying date-range filters at query time.
Additional error patterns emerged beyond temporal retrieval gaps. In one case, the gold answer contradicted the clinical documentation, and the model correctly identified the documented condition, indicating a labeling error in the benchmark rather than a system failure. In another case, patient education materials, which are generated regardless of whether the patient has the condition, appeared in the retrieved results, leading the model to incorrectly infer a diagnosis. This failure mode can be mitigated by using metadata filters to restrict retrieval to clinically substantive note types such as progress notes.

## System cost and performance

Using qwen3-embedding-0.6B with 300-token chunks and 50-token overlap, we produced 484 million vectors from 166 million notes. Table 1 summarizes infrastructure costs. The one-time index build cost was USD 891. Monthly operational costs totaled USD 4,021, comprising vector database (Vector Search) capacity and serving (USD 3,420), key-value store (Bigtable) serving and storage (USD 593), and networking (USD 7). These costs are independent of query volume within the provisioned capacity envelope of the deployed infrastructure. Embedding computation was provided at no cost through the TRC program; at on-demand pricing this would represent a substantial additional one-time cost. The web application runs on user-provisioned infrastructure.

To contextualize the storage-optimized pricing, we deployed an in-memory index containing 39.3 million vectors (8% of the full corpus), which alone cost approximately USD 8,000 per month, twice the full storage-optimized deployment. Based on consultation with Google Cloud engineers, a single-replica in memory deployment of the full corpus would cost an estimated USD 96,000 per month (costs scale in discrete units due to shard-based deployment in vector search).

After building the full index, we reran evaluation on the CHOP_MCQA_v0.5 dataset. Accuracy was 94.6% (95% Wilson CI: 91.6%--96.6%), a 0.9 percentage point decrease from the reduced benchmarking index. The confidence intervals overlap substantially, indicating no statistically significant degradation at scale.

We also assessed query latency by decomposing the end-to-end pipeline into three stages (Supplementary Figure 2). At single-user concurrency, median latency was 636 ms: 394 ms for query embedding on CPU, 237 ms for vector search, and 5 ms for key-value lookup. Because each user operates in an independent container, embedding latency scales horizontally. To assess how the shared index performed under concurrent query load, we benchmarked vector search latency at concurrency levels from 1 to 80 using pre-computed query embeddings. Because embeddings were supplied pre-computed, these measurements reflect vector search latency in isolation and do not represent full end-to-end query latency, which additionally includes query embedding time (394 ms on CPU) and key-value lookup (5 ms); single-user end-to-end latency is reported separately above. Median vector search latency increased from 237 ms with a single concurrent query to 451 ms at 20 concurrent queries, indicating that latency increased with load but remained under one second at moderate concurrency (Supplementary Figure 2). These results indicate that the shared index can support multiple simultaneous users with sub-second retrieval latency for the vector search stage, although search latency increases as concurrent demand rises.

## Direct retrieval evaluation

Our model-selection benchmark used end-to-end question-answering accuracy as a proxy for retrieval quality, a measure that cannot separate the contribution of retrieval from that of the language model's reasoning. To measure retrieval directly, we obtained human relevance judgments of the retrieved chunks for the selected production configuration (qwen3-embedding-0.6B with 300-token chunks); this labor-intensive annotation was not repeated for the other embedding models and chunk sizes. Across all 334 questions, at least one answering chunk appeared among the top 20 retrieved results for 87.4% (95% Wilson CI 83.4–90.5%), among the top 10 for 82.0% (77.6–85.8%), and among the top 5 for 73.7% (68.7–78.1%). An answering chunk was ranked first for 39.2% of questions (34.1–44.6%), corresponding to a mean reciprocal rank of 0.54, and mean precision@10 was 0.42. Inter-rater reliability was high (Cohen's $\kappa = 0.84$).

To interpret the relationship between retrieval and end-to-end accuracy, we categorized each question by whether its correct answer asserted a finding was present (positive, n=236) or absent/normal (negative, n=98) and compared performance with and without retrieval. Retrieval improved accuracy for both. For positive questions, accuracy rose from 22.0% without retrieval to 94.9% with it (Hit@20 97.9%), confirming that the system surfaces the answering evidence and the model relies on it. For negative questions, accuracy rose from 71.4% to 93.9%; here retrieval acted chiefly by suppressing hallucinated findings. Of the 25 negative questions retrieval corrected, 21 were cases in which the context-free model asserted a condition the patient did not have, which the retrieved notes overrode. Retrieval thus contributes through two distinct mechanisms: supplying answering evidence for findings and suppressing unsupported assertions for absences.

**Clinical utility evaluation**

To assess clinical utility, five clinicians completed three abstraction tasks inspired by active research projects: documenting a patient's genetic diagnosis as free text, later harmonized to categorical values (task 1), recording age in months at first documented seizure (task 2), and determining cohort eligibility for ballet-related foot injuries (task 3). Semantic search consistently reduced abstraction time across all three abstraction tasks (Figure 4). For genetic conditions (task 1), median time decreased from 40 seconds to 30 seconds (24.1% reduction, $p = 0.013$, $r = 0.29$). For seizure documentation (task 2), median time decreased from 67 seconds to 32 seconds (51.9% reduction, $p < 0.001$, $r = 0.47$). For cohort identification of ballet-related foot injuries (task 3), median time decreased from 260 seconds to 28 seconds (89.4% reduction, $p < 0.001$, $r = 0.83$). Figure 4 summarizes per-patient abstraction times by task and search method.

To assess whether the two search methods yield equivalent extracted information, inter-rater agreement was computed overall and separately for within-method pairs (EHR vs EHR) and cross-method pairs (EHR vs Semantic) for tasks 1 and 2. For genetic conditions, overall inter-rater agreement was high (Fleiss' $\kappa = 0.945$), and agreement was similar when stratified by method (within-method Cohen's $\kappa = 0.957$; cross-method Cohen's $\kappa = 0.925$). For seizure age, overall agreement was likewise high (Krippendorff's $\alpha = 0.950$), with comparable results by method (within-method $\alpha = 0.977$; cross-method $\alpha = 0.950$). For both tasks, the difference between within-method and cross-method agreement was not statistically significant (genetic conditions: bootstrap $p = 0.47$; seizure age: bootstrap $p = 0.37$). These results suggest that semantic search maintains abstraction agreement at a level comparable to traditional EHR abstraction, while requiring significantly less time.

Inter-rater agreement was not assessed for task 3 because the two methods did not operate on a shared patient list, precluding direct agreement comparison. Task 3 was intentionally asymmetric because the two methods work in fundamentally different ways: the substring-query arm searches within a fixed set of keyword matches, whereas semantic search must find eligible patients from the full corpus with no predefined list. The time savings therefore reflect the complete discovery task, not the same work done two ways. To assess whether faster discovery came at a cost to accuracy, we retrospectively adjudicated the eligibility of every uniquely selected patient, blinded to search arm. All selected patients were confirmed eligible in both arms (semantic search 30/30, precision 100%, 95% Wilson CI 89–100%; substring-query EHR review 22/22, precision 100%, 95% Wilson CI 85–100%).

**Cross-patient genetic disease retrieval**

To assess health-system-scale retrieval more broadly, we tested the system's ability to retrieve cohorts of patients with molecularly confirmed genetic diagnoses. For each condition, we created a single natural-language query designed to identify matching patients (Supplementary Table 3). Across 71 cohorts (812 confirmed patients), the query with no structured filter recovered 90.5% of confirmed patients within the top 300 unique candidates, 97.4% within 1,000, and 98.5% within 10,000 (patient-weighted; Table 2, Figure 5).

To compare this performance with traditional approaches, we performed an ICD-10 code-based search. Structured retrieval was fundamentally bounded. Even counting a patient as retrievable if they carried any relevant diagnosis code at any depth, ICD-based search reached only 75.0% of confirmed patients (patient-weighted recall ceiling), fewer than semantic search recovered within the first 300 candidates alone. At matched retrieval depths, semantic search recovered 90.5% versus 58.6% (ICD) at k=300 and 97.4% versus 69.9% at k=1,000 (Table 2). Reviewing exactly as many candidates as there were confirmed patients (k = cohort size), semantic search recovered 27.6% versus 20.3% for the structured comparator. The structured ceiling arises because a molecularly confirmed patient without a corresponding coded encounter is unretrievable by code at any depth. For 29 of the 71 conditions, no ICD-10-CM code names the syndrome. Even among conditions with a specific code, many confirmed patients—frequently presymptomatic relatives identified through cascade testing—had never generated a coded encounter for the phenotype. Augmenting the comparator with family-history codes did not materially raise its ceiling (75.4%).

Three conditions illustrate the range (Figure 5). For *ENG*-associated hereditary hemorrhagic telangiectasia (n=18), a symptomatic and consistently coded condition, the two methods performed comparably: both recovered all confirmed patients, structured search within 300 candidates and semantic search within 1,000. For *MYBPC3*-related hypertrophic cardiomyopathy (n=52), a specific diagnosis code exists, yet only 42% of confirmed patients had ever been coded for cardiomyopathy, capping structured recall at that level; semantic search recovered 96.2% within 300 candidates and 100% within 1,000. For *BRCA1*-associated hereditary breast and ovarian cancer (n=19), no specific ICD-10-CM code exists and the best genetic-susceptibility surrogate reached only 26% of confirmed patients, whereas semantic search recovered 89.5%.

**Discussion**

We have demonstrated that semantic search over the entirety of a health system's clinical notes is technically feasible and operationally sustainable. The system indexes 166 million notes as 484 million vectors, returns results in under one second, and costs approximately USD 4,000 per month to operate regardless of query volume. We achieved a high accuracy on a physician-authored clinical question-answering benchmark, and inter-rater agreement between semantic search and traditional EHR abstraction was comparable, indicating that the system maintains high retrieval quality. To our knowledge, this is the first report of semantic retrieval deployed across an entire health system EHR.

Embedding models, vector databases, and approximate nearest neighbor algorithms are established technologies. Our contribution is the demonstration that they compose into a working system at this scale and cost, and support real-world chart abstraction workflows. Several design decisions were critical. Separating vector storage (in the index) from full-text storage (in the key-value store) reduced costs by avoiding the expense of storing note content alongside high-dimensional vectors. Selecting a storage-optimized index over an in-memory configuration reduced monthly hosting costs, with a subset comparison showing that 8% of the corpus in memory costs more than 100% on storage-optimized disk. The latency trade-off remained within sub-second bounds acceptable for interactive use.

A further design decision with implications for both usability and retrieval quality was the integration of metadata filters, which serve a dual purpose. They enable users to scope queries by patient, time, specialty, note type, and other attributes, a natural extension of how clinicians think about patient information. They also mitigate a known property of dense retrieval: similarity scoring favors short, narrowly focused documents whose embeddings are tightly aligned with queries. Filtering by note category allows users to prioritize clinically substantive documents such as progress notes and discharge summaries over routing notes or patient education documentation.

Our evaluation of clinical utility demonstrated that the system reduced the time necessary for chart review tasks while maintaining a high level of accuracy in the abstracted data, consistent with prior work showing that traditional chart review is time-intensive and difficult to scale[27,28]. The system has already been adopted by several principal investigators and their research teams and is continuing to see broader uptake by additional researchers across CHOP. Initial user interactions and feedback revealed two observations. First, users valued the ability to query the entire patient population simultaneously, with sub-second response times contrasting sharply with conventional one-patient-at-a-time chart review. For rare disease research or cases where keyword search produces many false positives and fails on negations and synonyms, semantic search showed promise for improving cohort identification, particularly given that keyword-based approaches have demonstrated poor specificity and lack of semantic understanding[29–31]. Second, users noted that relevant information was surfaced from overlooked note types, such as nurse triage and telephone encounters, that are often skipped in traditional chart review but contain critical clinical details. Formalizing these observations through longitudinal user studies and measuring their impact on research outcomes remains important for future work.

Beyond supporting human chart review, high-quality retrieval also benefits the automated language-model applications the system is designed to ground, where a frequently raised concern is the fabrication ("hallucination") of patient-specific facts. Our comparison of the system with and without retrieval speaks directly to this concern: absent the patient's notes, the model asserted conditions that were not present, whereas grounding it in the retrieved record suppressed the large majority of these hallucinated findings. This demonstrates that retrieval serves not only to supply answering evidence but to constrain the model to what is actually documented.

Retrieval performance varied substantially across questions in ways that tracked how and where the answering evidence was documented in each patient's record, rather than the embedding model, which was held constant in this analysis; characterizing this question–documentation interaction—why some clinical questions are intrinsically more retrievable than others—is an important direction for future work.

The system is designed as an infrastructure rather than a standalone application. The retrieved content serves as grounding evidence for downstream tasks including retrieval-augmented generation, and the index can support programmatic access for automated workflows. The governance framework, including per-project containerized deployments, allowlist-based access control, query audit logging, and a standardized IRB protocol attachment provides the institutional scaffolding needed for broad adoption.

Beyond within-chart abstraction, semantic search supported patient discovery at the scale of the entire health system. Against a molecularly defined reference standard of genetic-disease cohorts, a single natural-language query recovered the large majority of confirmed patients from the full corpus, substantially more than a structured query built from diagnosis codes. The difference is structural rather than incidental: code-based retrieval can only recover patients who carry a relevant code. However, for many genetic conditions no specific diagnosis code exists. Even where one does, molecular confirmation frequently precedes or occurs without a coded encounter, as when presymptomatic relatives are identified through cascade testing. Because the confirming evidence resides in the free text, semantic search recovers these patients regardless of coding status. Where a condition was both specifically codeable and clinically manifest, structured retrieval performed comparably. Moreover, each retrieved patient is returned together with the specific note passage that matched the query, giving a reviewer the evidence needed to adjudicate eligibility directly. Combined with inexpensive large language model triage of the ranked candidates, this makes even large candidate sets practical to screen and positions the system to assemble cohorts, including previously unascertained patients, that structured queries cannot.

**Limitations and future work**

The primary limitation of this study is that the evaluation is limited to a single pediatric academic medical center deployed on Google Cloud Platform; documentation practices, optimal system configurations, and infrastructure choices may differ across institutions and cloud providers. Furthermore, the system indexes only text and does not support multimodal data such as images or tables; extending the system to support multimodal inputs would provide richer context for complex clinical tasks. Additionally, the chunking strategy relies on fixed-size windows with linguistic boundary heuristics and does not account for semantic coherence; developing content-aware semantic chunking methods that segment notes based on clinical concept boundaries is an important direction for future work. The cohort-retrieval evaluation reports recall against an institution-specific registry that is not an exhaustive enumeration of confirmed patients; precision could therefore not be estimated, and prospective assessment of patients surfaced beyond the registry remains for future work.

Regarding portability, it is important to distinguish architectural decisions that are vendor-specific from those that are provider-agnostic. The production deployment relies on Google Cloud Vector Search for query-time metadata filtering, the ability to restrict nearest-neighbor retrieval by structured attributes such as patient identifier, note type, and date range at search time, rather than post-hoc. This capability is not universally available across managed vector database services, and porting the system to other providers would require evaluating equivalent functionality or adapting the retrieval pipeline accordingly. A porting guide documenting these considerations is included in the reference implementation repository. In contrast, the remaining core architectural decisions, including the choice of embedding model, the chunk-size and overlap strategy, and the separation of vector storage from full-text retrieval, are provider-agnostic and have direct equivalents across major cloud platforms and open-source vector databases. Validation at institutions using different EHR vendors, cloud providers, or patient populations (including adult health systems) is an important direction for future work.

## Conclusion

Health-system scale semantic search over unstructured clinical notes is technically feasible with sub-second latency and low operational cost. By combining a high-performance open-source embedding model, a storage-optimized vector index, and a low-latency key-value store, we indexed 484 million vectors from 166 million clinical notes. Importantly, this work moves semantic search from a proof-of-concept task to shared institutional infrastructure. Through a researcher-facing interface that supports natural-language queries and fine-grained filters, investigators can perform chart review and cohort discovery with accuracy comparable to traditional EHR review methods. The system provides institutional infrastructure for interactive search, automated abstraction, and LLM-powered clinical reasoning across the health system.

## Data availability

The data used in this study were accessed under a waiver of HIPAA authorization and consent (see Methods) and cannot be transferred outside the covered entity. Researchers at other institutions may conduct analogous studies by deploying the reference implementation (see Code Availability) against their own institutional data under a locally approved protocol.

## Code availability

A portable reference implementation of the system described in this paper is publicly available at https://github.com/Ian-Campbell-Lab/clinical-semantic-search. The repository includes the core search pipeline (chunking, embedding, vector search, metadata retrieval), the web-based search interface, benchmarking tools for latency and retrieval quality, and reference implementations of all cloud service integrations. All institution-specific configuration is externalized to environment variables, and a porting guide provides step-by-step instructions for adapting the system to other health systems, cloud providers, and EHR vendors.

## Author contributions

F.W.M., S.M., and I.M.C. conceived the study and developed the overall methodology. S.D., P.D., S.K.B., S.H., K.M., R.W.G., S.M.H., and J.M.M. provided methodological guidance and contributed to system architecture development. A.B.R., I.R.R., and I.M.C. created the benchmarking dataset. F.W.M. and A.L. performed the experiments and data analysis. A.L., I.R.R., B.H.C., S.C., R.W.G, and I.M.C. conducted the clinical utility evaluation. F.W.M. wrote the original draft. A.L. and I.M.C. reviewed and revised the manuscript. R.W.G, H.M.G., and I.M.C. supervised the project. All authors reviewed and approved the final manuscript.

**Acknowledgements**

This project received complimentary computational support through the Google TPU Research Cloud (TRC). I.M.C was supported by grant K08-HD111688 and S.C by grant T32-HD060550 from the Eunice Kennedy Shriver National Institute of Child Health and Human Development.

**Competing interests**

S.D. is an employee of Google Cloud LLC. I.M.C. reports research support from Google Cloud LLC. Google Cloud LLC had no oversight or control over the experimental design or publication of the work. All other authors declare no competing interests.

## Figures

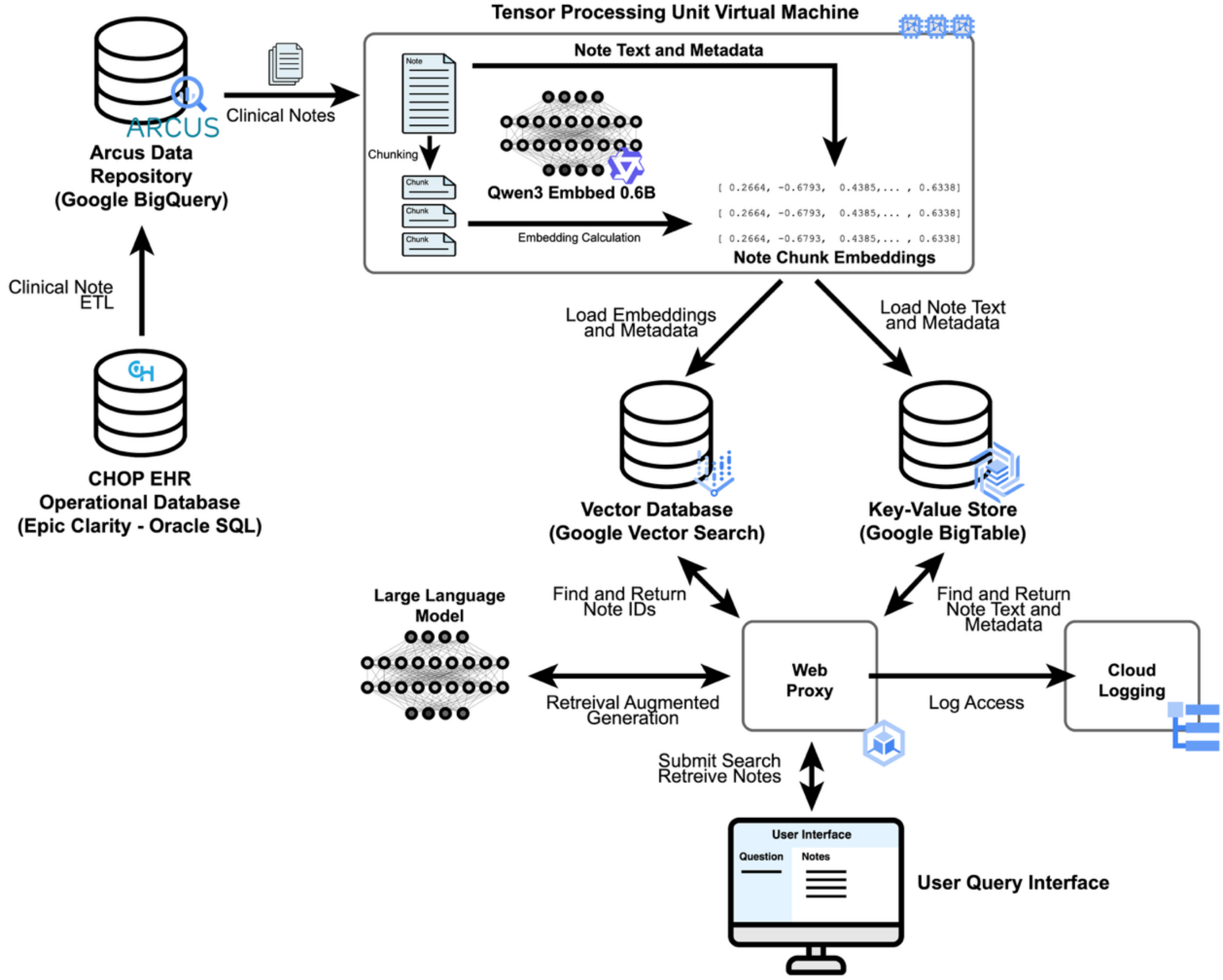

**Figure 1:** System architecture for health system–scale semantic search. Clinical notes are extracted from the CHOP EHR database, chunked into 300-token chunks with 50-token overlap and embedded using qwen3-embedding-0.6B. Chunk embeddings are stored in a vector database, while the full note text and metadata are stored in a low-latency key-value store (chosen for cost efficiency since BigTable offers lower-cost storage compared with the vector database). User queries and search results are logged for auditing and usage analysis. All aspects occur within a secure HIPAA-compliant environment.

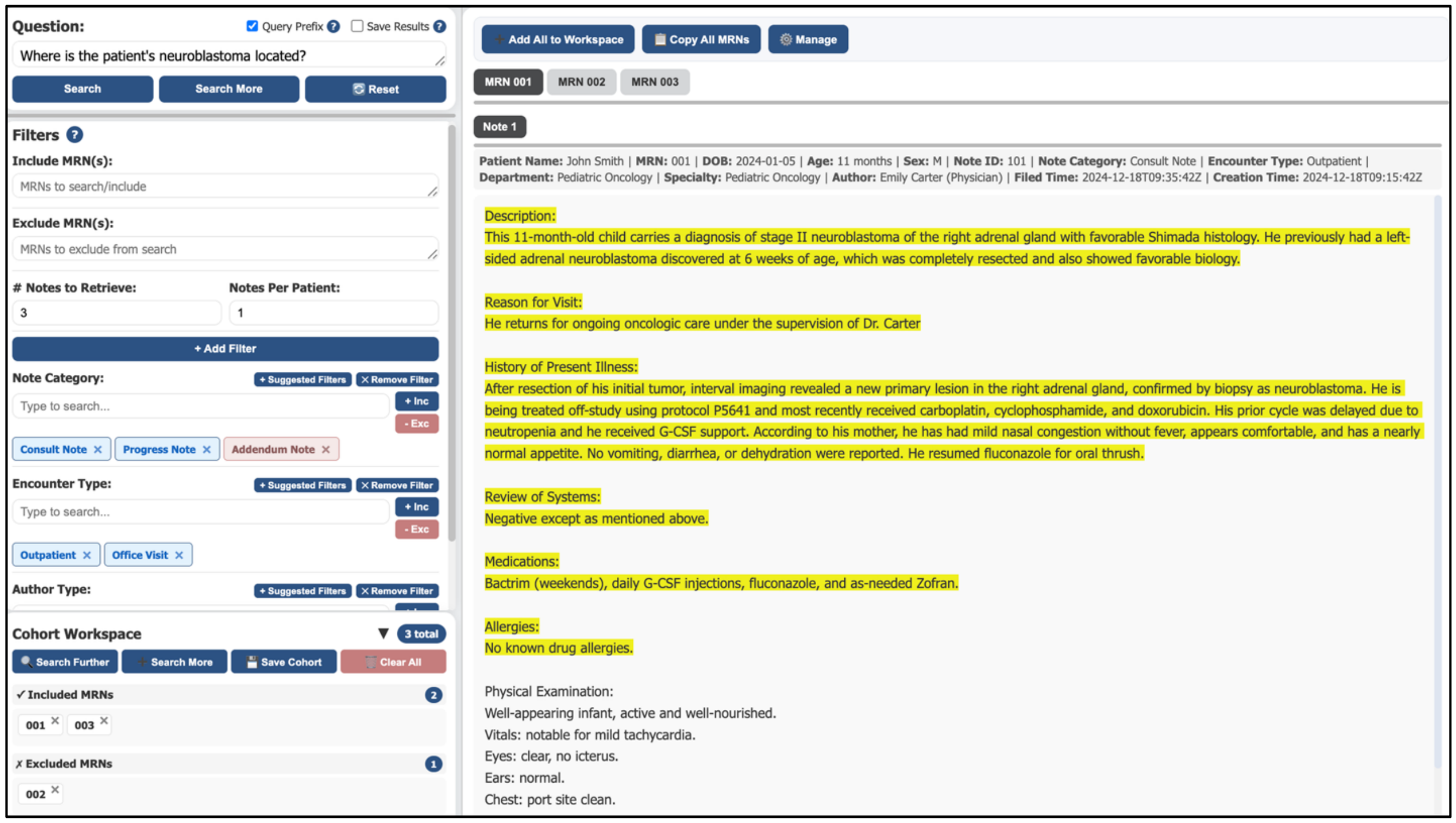

**Figure 2:** Screenshot of the semantic search user Interface. The interface enables users to query the semantic search system using natural language and apply advanced filters to refine results. **Left panel:** Users enter a question and select filters such as patient identifier (MRN), number of notes to retrieve, note category, encounter type, and so on to customize the search. Additional cohort-building tools allow users to include or exclude patients based on retrieved evidence. **Right panel:** Search results are displayed and grouped by patient. Highlighted text indicates a 300-token chunk that likely answers the user's query within the full clinical note. Users can iteratively refine queries, review retrieved evidence, and curate patient cohorts using the Manage and Cohort Workspace features, enabling efficient cohort construction without requiring technical expertise. All data in this screenshot is synthetic.

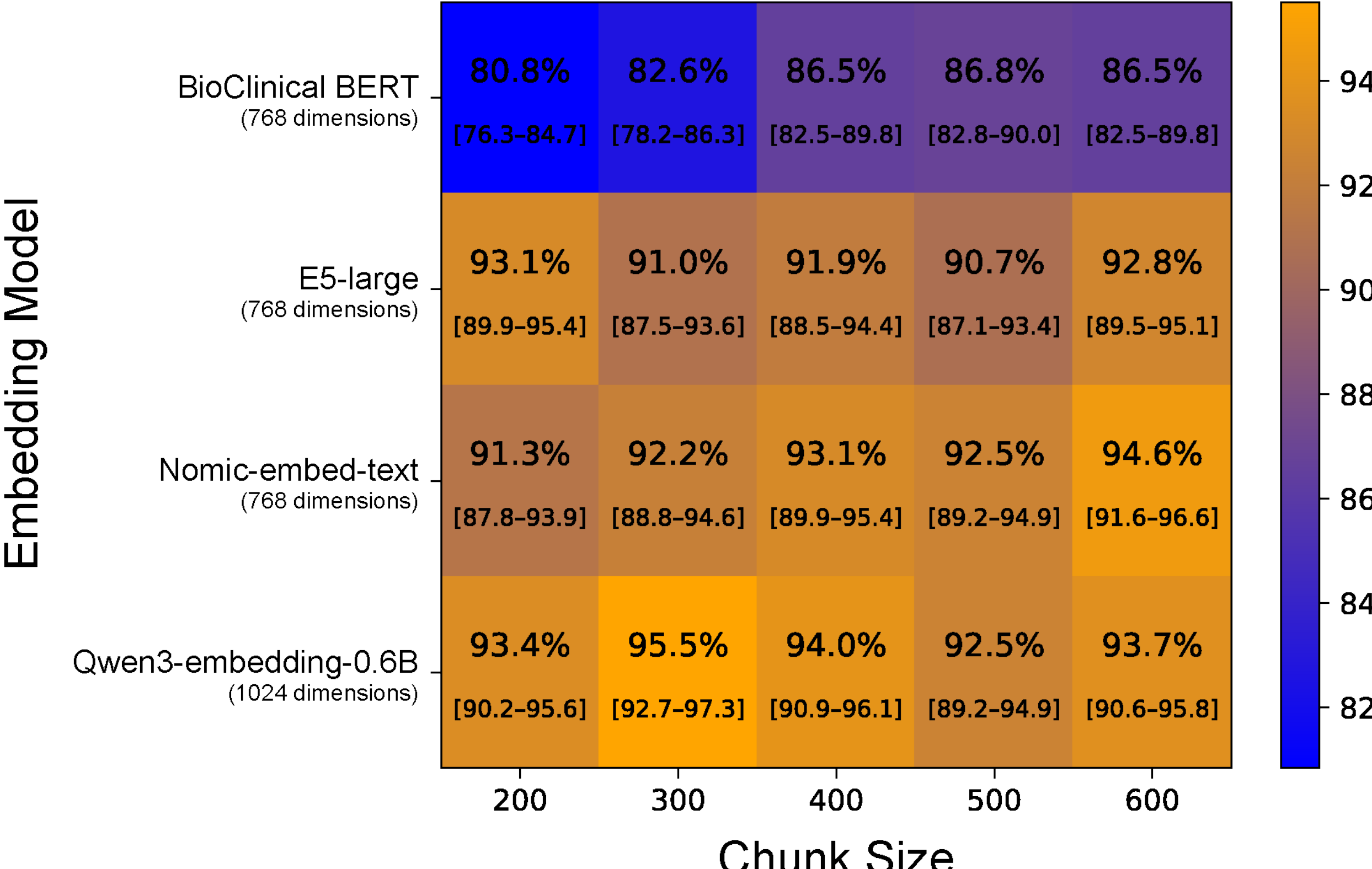

**Figure 3:** Accuracy evaluation of embedding models and chunking strategies on the CHOP_MCQA_v0.5 benchmark. Qwen3-embedding-0.6B with a 300-token chunk size achieved the highest accuracy of 95.5%. These experiments were run on a reduced index containing only the notes for patients included in the benchmarking dataset. Wilson score 95% confidence intervals are shown in brackets.

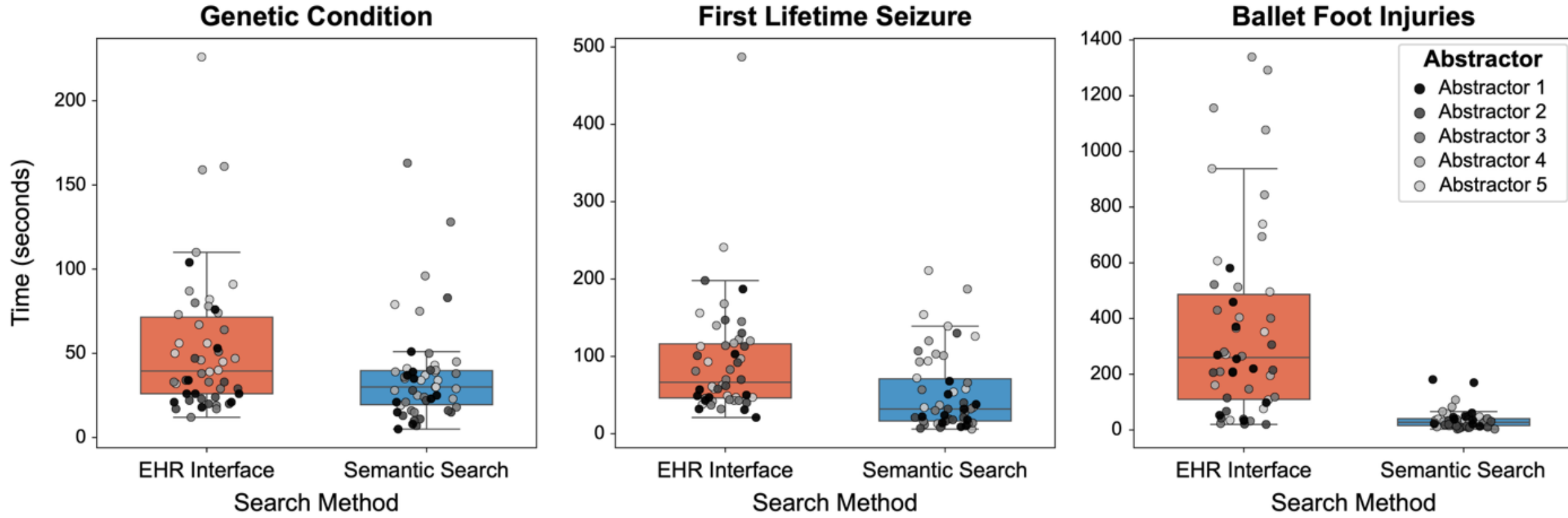

**Figure 4:** Comparison of abstraction time by search method across three clinician tasks. **Left:** identification of documented genetic conditions (task 1). **Middle:** documentation of age at first lifetime seizure (task 2). **Right:** cohort generation for ballet-related foot injuries (task 3). Each panel shows boxplots of per-patient time-to-completion (seconds) for the EHR keyword-search condition (left box in each panel) versus the semantic-search condition (right box), with overlaid individual points for each of the five abstractors (different marker shades). Across all three tasks, median completion times significantly decreased: for task 1 from 40 to 30 seconds (24.1% reduction, $p = 0.013$ ), for task 2 from 67 to 32 seconds (51.9% reduction, $p < 0.001$), and for task 3 from 260 to 28 seconds (89.4% reduction, $p < 0.001$).

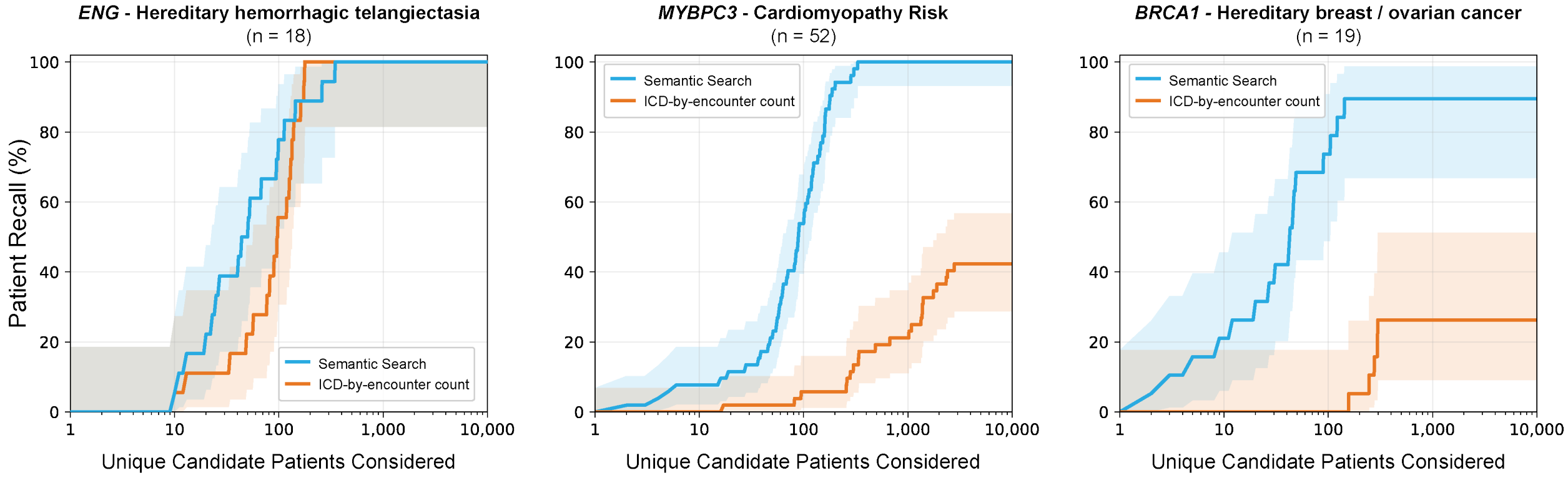

**Figure 5:** Corpus-wide patient recall as a function of the number of unique candidate patients reviewed, for semantic search (natural-language query) versus structured ICD-by-encounter-count retrieval, across three conditions spanning the range of coding completeness. **Left:** *ENG*-associated hereditary hemorrhagic telangiectasia (n=18), a symptomatic, consistently coded condition, where both methods recover all confirmed patients. **Middle:** *MYBPC3*-related hypertrophic cardiomyopathy (n=52), which has a specific diagnosis code, yet most confirmed patients, frequently presymptomatic relatives, were never coded for the phenotype. Right: *BRCA1*-associated hereditary breast and ovarian cancer (n=19), for which no specific ICD-10-CM code exists. Semantic recall (blue) rises with retrieval depth; structured recall (orange) plateaus at its ceiling, the fraction of confirmed patients carrying any relevant diagnosis code (dashed line), because molecularly confirmed patients without a coded encounter are unretrievable by code at any depth. Shaded bands are 95% Clopper-Pearson confidence intervals on recall.

**Tables**

**Table 1:** Infrastructure Costs

| **Infrastructure Element** | **Monthly Cost** |
|---|---|
| Vector Search Capacity | $3420 |
| Big Table Server | $535 |
| Big Table Solid State Drive | $58 |
| Networking | $7 |
| Logging | $1 |
| Total | $ 4,021 |

**Table 2:** Patient-level Recall of Genetic Conditions

| **Method** | **Recall@CohortSize** | **Recall@300** | **Recall@1000** | **Recall@Ceiling** |
|---|---|---|---|---|
| Semantic search | 27.6% | 90.5% | 97.4% | 98.5% |
| ICD-by-encounter-count | 20.3% | 58.6% | 69.9% | 75.0% |

Recall@CohortSize is recall when the number of candidates reviewed equals each condition's cohort size; Recall@Ceiling is recall at a depth of 10,000 candidates for semantic search, and the fraction of confirmed patients carrying any relevant ICD code (the structural maximum) for ICD-by-encounter-count retrieval.

**Supplementary Information**

**Supplementary Table 1**: Filterable index metadata

| **Filter** | **Description** |
|---|---|
| Medical record number (MRN) | Retrieve notes for one or more specific patients |
| Age | Limit results to specific age range of the patient on the date the note was written |
| Note category | Limit or exclude results based on specific categories of the note (e.g., progress notes, discharge summaries, consult notes) |
| Encounter type | Limit or exclude results based on the type of encounter in which the note was written (e.g., hospital encounter, office visits, telephone encounters) |
| Department | Limit or exclude results based on the department in which the note was written (i.e. outpatient endocrinology, particular primary care site) |
| Specialty | Limit or exclude results based on specialty (e.g., cardiology, emergency medicine, genetics) |
| Author type | Limit or exclude results based on particular author roles (e.g., physicians, fellow, therapist, nurse) |
| Author name | Limit or exclude results based on particular named authors |
| Date | Limit or exclude results based on the date and time the note was written |

**Supplementary Table 2**: CHOP_MCQA_v0.5 benchmarking example questions used to evaluate system performance.

| **Category** | **Example** |
|---|---|
| Cardiology | Which of the following best describes the patient's heart anatomy at birth? |
| Rheumatology | Which of the following best describes the patient's rheumatologic disease history? |
| Hematology | Which of the following best describes the patient's anemia history? |
| Genetics | Which of the following best describes whether the patient has been diagnosed with a genetic condition? |
| Nephrology | Which of the following best describes the patient's renal disease history? |
| Hospitalization | Which of the following best describes the primary reason for the patient's hospitalization at CHOP in [DATE/ YEAR]? |
| Devices | Which of the following best describes why the patient's medical device was removed intraoperatively in [DATE/ YEAR]? |
| Endocrinology | Which of the following best describes history of how the patient's diabetes was diagnosed? |
| Allergy/ Immunology | What was the trigger of the patient's episode of anaphylaxis treated in the ED in [DATE/ YEAR]? |
| Oncology | What oncologic diagnoses does this patient have, if any? |
| Infectious Diseases | Which organism was responsible for the patient's bloodstream infection in [DATE / YEAR]? |
| Social Work | Referral was made to which community resource to address the patient's [ISSUE]? |
| Critical Care | Which of the following best describes the cause of the patient's acute kidney injury during their PICU stay in [DATE/ YEAR]? |
| Child protection | Which of the following best describes the areas of this patient's body have had physical exam findings concerning for bruising? |
| Neonatology | Which of the following reasons best describes why the patient was admitted to the CHOP NICU in [DATE/ YEAR]? |

**Supplementary Table 3:** Per-condition detail for the cross-patient cohort-retrieval evaluation. For each of the 71 Genomic Indicator conditions: gene, indicator label, confirmed cohort size, the natural-language query issued, the ICD-10-CM code(s) used for the structured comparator (with any family-history code and whether a specific code exists), and per-condition recall for semantic search and ICD-by-encounter-count retrieval.

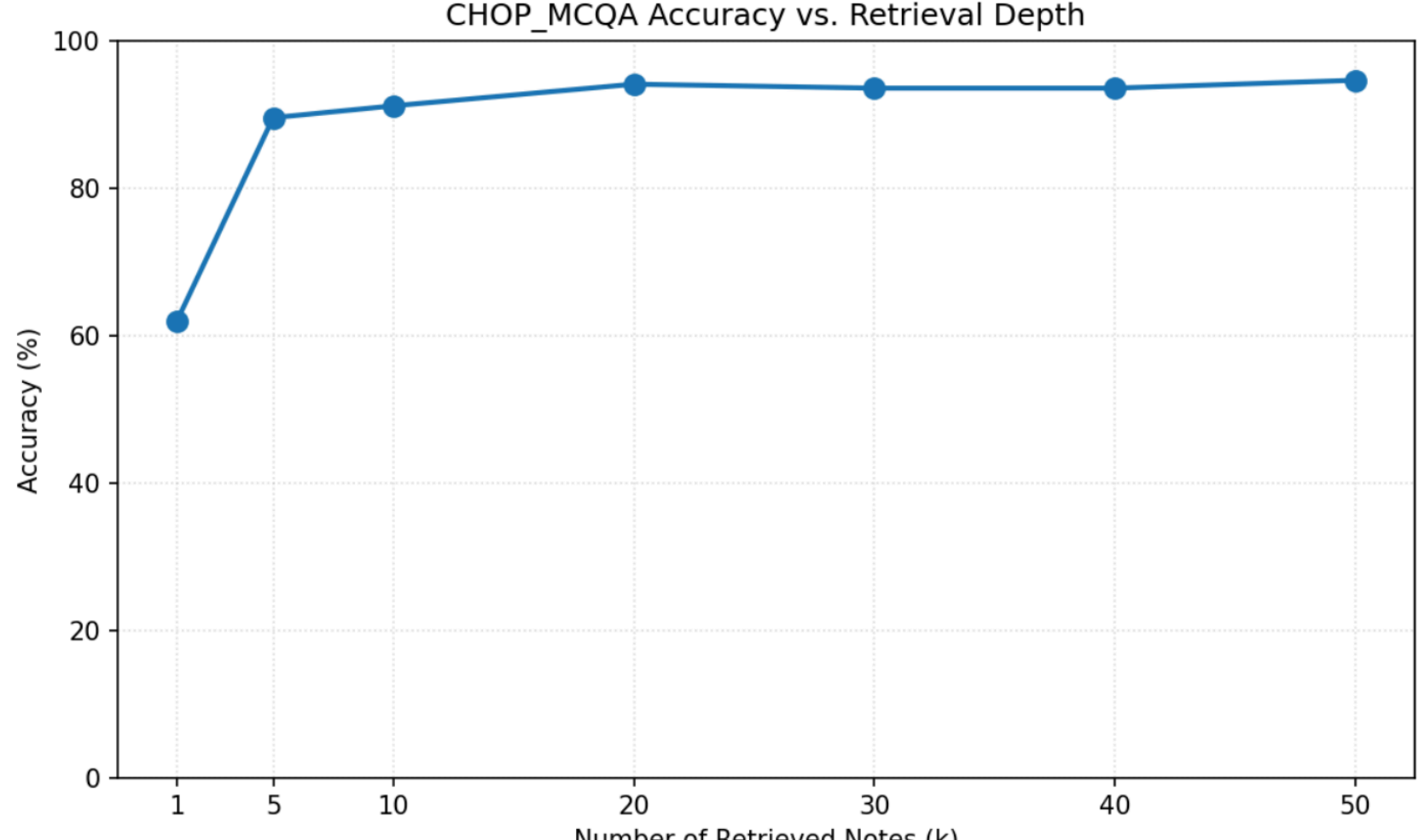

**Supplementary Figure 1**: Accuracy (%) on the CHOP_MCQA_v0.5 benchmark dataset measured after retrieving different numbers of notes (k = 1, 5, 10, 20, 30, 40, 50) shows improvements as retrieval depth increases and peaking near k=20. Performance plateaus after ~20 retrieved notes, indicating diminishing returns from deeper retrieval.

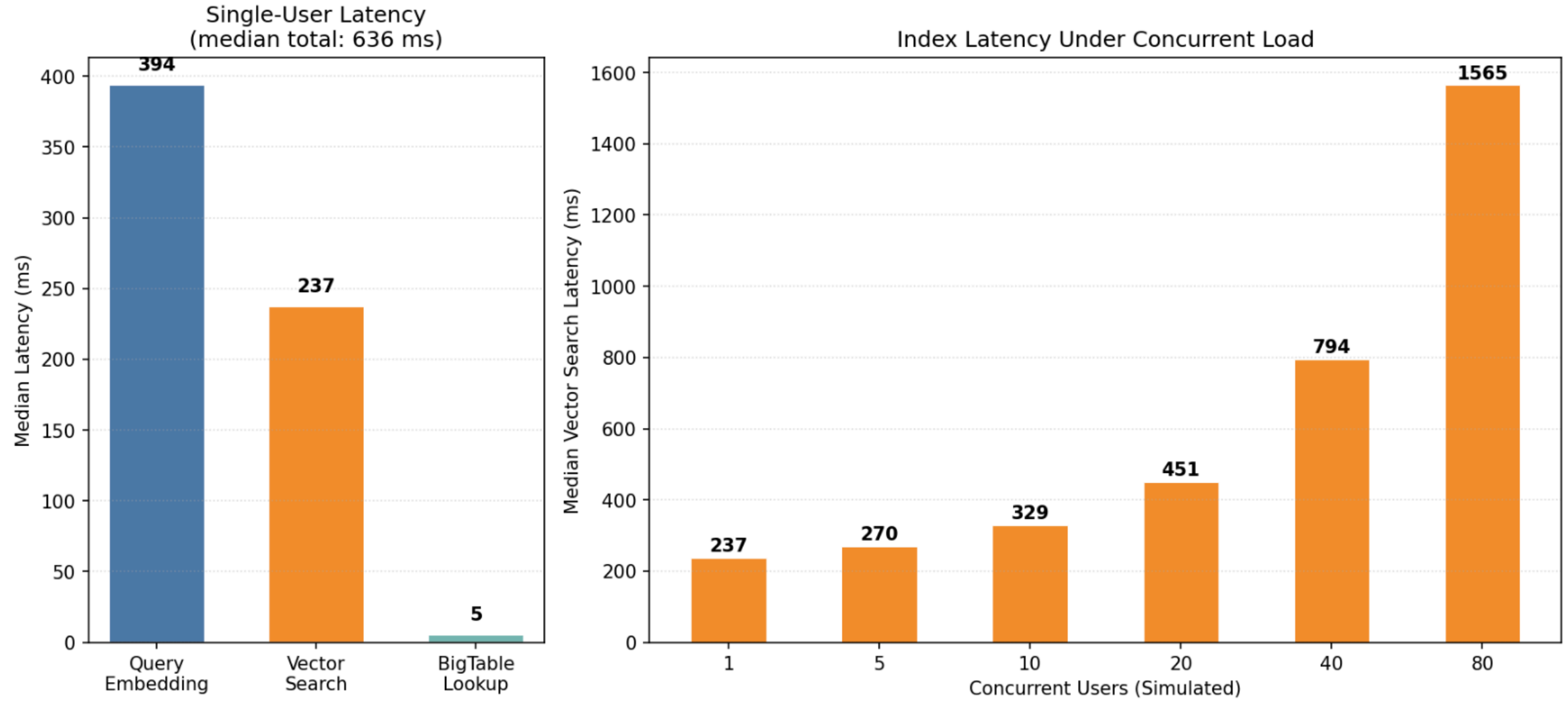

**Supplementary Figure 2**: Single-user end-to-end query latency and shared-index vector search performance under concurrency. **Left:** Breakdown of median single-user latency (total 636 ms) into three stages : query embedding on CPU (394 ms), vector search (237 ms), and Bigtable lookup (5 ms). **Right:** Median vector-search latency for a shared index measured with pre-computed embeddings (vector search stage only, excluding query embedding and key-value lookup) at simulated concurrent users (1, 5, 10, 20, 40, 80). Median search latency rises from 237 ms at concurrency 1 to 451 ms at 20 and 1,565 ms at 80, but remains under one second for multi-user workloads.